\documentclass[submission, fleqn, Phys]{SciPost}
\pdfoutput=1
\usepackage{amsmath,amssymb}
\usepackage[binary-units]{siunitx}

\newcommand{\sherpa}{Sherpa}
\newcommand{\madgraph}{MadGraph}
\newcommand{\amegic}{Amegic}
\newcommand{\comix}{Comix}
\newcommand{\rambo}{Rambo}

\newcommand{\tess}{Tess}
\newcommand{\newalgo}{BlockGen}
\newcommand{\newalgolc}{\newalgo-LC}
\newcommand{\newalgoco}{\newalgo-CO}
\newcommand{\newalgocosummed}{\newalgoco$_\Sigma$}
\newcommand{\newalgocosampled}{\newalgoco$_\text{MC}$}
\newcommand{\newalgocdsampled}{\newalgo-CD$_\text{MC}$}

\begin{document}

\begin{flushright}
  FERMILAB-PUB-21-263-T
\end{flushright}
\begin{center}{\Large \textbf{
  Many-gluon tree amplitudes on modern GPUs\\[1mm]
  A case study for novel event generators
}}\end{center}
\begin{center}
E. Bothmann\textsuperscript{1*},
W. Giele\textsuperscript{2},
S. H{\"o}che\textsuperscript{2},
J. Isaacson\textsuperscript{2},
M. Knobbe\textsuperscript{1}
\end{center}

\begin{center}
{\bf 1} Institut für Theoretische Physik, G{\"o}ttingen, Germany\\
{\bf 2} Fermi National Accelerator Laboratory, Batavia, USA\\
$^*$ enrico.bothmann@uni-goettingen.de
\end{center}

%\begin{center}
%\today
%\end{center}
% For convenience during refereeing: line numbers
%\linenumbers

\section*{Abstract}
{\bf
  The compute efficiency of Monte-Carlo event generators for the
  Large Hadron Collider is expected to become a major bottleneck for
  simulations in the high-luminosity phase. Aiming at the development
  of a full-fledged generator for modern GPUs, we study the performance
  of various recursive strategies to compute multi-gluon tree-level amplitudes.
  We investigate the scaling of the algorithms on both CPU and GPU hardware.
  Finally, we provide practical recommendations as well as baseline implementations
  for the development of future simulation programs. The GPU implementations can be
  found at: \url{https://www.gitlab.com/ebothmann/blockgen-archive}.
}

\section{Introduction}
\label{sec:intro}
The success of high-energy physics experiments at particle colliders
critically depends on a detailed computer simulation of the collisions.
So-called event generators link theory and experiment 
through an event-by-event representation of the observable 
final-state particles by means of Monte-Carlo integration methods.
A complete event generator consists of modules, which encapsulate
the physics at various energy scales~\cite{Webber:1986mc,Buckley:2011ms}.
Interactions at the highest scales, which are usually regarded as
the most relevant to probing our fundamental understanding of nature,
are simulated by matrix element generators.
In a community effort, these programs have been nearly fully automated.
They can be used for computations at tree-level~\cite{
  Kanaki:2000ey,Papadopoulos:2000tt,Krauss:2001iv,
  Gleisberg:2008fv,Maltoni:2002qb,Alwall:2007st,Alwall:2011uj,
  Mangano:2002ea},
and at one-loop level~\cite{Ossola:2007ax,Gleisberg:2007md,
  Berger:2008sj,Bevilacqua:2011xh,Cascioli:2011va,
  Cullen:2014yla,Alwall:2014hca,Actis:2016mpe}
in perturbation theory and can handle
a large class of fundamental and effective theories~\cite{
  Christensen:2009jx,Degrande:2011ua,Staub:2013tta}.

Such flexibility typically comes at the cost of compute efficiency.
This is currently a major problem, and will be prohibitive for the
high-luminosity phase of the LHC, where the projected reach in terms
of final-state multiplicities and large statistics data sets
poses new challenges
to event generator codes~\cite{Calafiura:2729668,HSFPhysicsEventGeneratorWG:2020gxw}.
Some event generator collaborations have therefore explored
massively parallel
computing techniques~\cite{Hoeche:2013zja,
Alves:2017she,Dainese:2019rgk}.
The complexity of existing generators, and the requirement to
continuously provide state-of-the-art simulations for LHC physics
have so far prevented the wider adoption of new computing paradigms,
such as GPUs. In the past, several conceptual studies have shown
substantial performance gains, both for tree-level computations~\cite{
  Hagiwara:2010,Hagiwara:2010a,Kanzaki:2011,Hagiwara:2013,Kanzaki:2010ym,
  Giele:2011,Wu:2019tsf,Grasseau:2015vfa,Carrazza:2020,Carrazza:2021zug}
and for one-loop level computations~\cite{Yuasa:2013qza,Borowka:2018goh}.
In addition, the evaluation of PDFs~\cite{Carrazza:2020qwu} and
the matrix element method~\cite{Schouten:2015,Grasseau:2015,Grasseau:2019sih}
have been explored on GPUs.
Nevertheless, a production-ready GPU-enabled event generator suitable for
experimental applications has not yet become available.
The construction of such a generator must begin with the selection
of the computational algorithm which has the best metrics in terms
of flexibility and efficiency.
This algorithm should also be easy to understand and implement,
such that portability questions can be discussed
not only by high-energy physicists,
but by a somewhat larger scientific computing
and computer science community.

Based on many previous perturbative QCD and Beyond Standard Model
applications, we argue that such an algorithm is given by the
Berends--Giele recursion~\cite{Berends:1987me,Badger:2012uz}.
Due to its inherently parallel nature, the recursion can easily be
thread-parallelized~\cite{Gleisberg:2008fv,Giele:2011}.
In addition, modern computer architectures with their
large memory and efficient memory access allow the trivial
parallelization of the entire event loop in the Monte-Carlo integration.
The computation of color factors needed for the assembly of full
matrix elements can either be carried out in a factorized form,
or embedded in the recursion itself~\cite{Duhr:2006iq,Giele:2010}.
We refer to the factorized option as a color-ordered Berends--Giele (COBG) recursion
and to the embedded one as a color-dressed Berends--Giele (CDBG) recursion.
From a computational point of view, both have advantages and disadvantages,
and which method is most suitable for a certain hardware depends critically
on the size, bandwidth and access pattern of fast memory.

In this paper, we therefore perform a comprehensive study of the
known variants of the Berends--Giele recursion and assess their strengths
and weaknesses on various computing platforms. We investigate the scaling
of computation time and memory usage with the number of external gluons
in all-gluon amplitudes,
and compare the performance of our new CPU and GPU codes to the existing
general-purpose matrix element generators \madgraph~\cite{Alwall:2011uj},
\amegic~\cite{Krauss:2001iv}, and \comix~\cite{Gleisberg:2008fv}.
We start from the thread-scalable Berends--Giele recursion~\cite{Giele:2010},
which uses the leading color approximation,
and extend it to full color and helicity summed amplitudes,
by using: naive sums, color sampling in the color-flow
basis~\cite{Maltoni:2002mq}, color dressing~\cite{Duhr:2006iq}
and the continuous color and helicity representation
of~\cite{Draggiotis:1998gr}.
For a realistic performance assessment of the algorithms,
we do not take advantage of symmetries
that are specific to all-gluon scattering,
and we only take into account algorithms
that generate strictly positive weights.

This paper is organized as follows. In Sec.~\ref{sec:theory} we
give a brief introduction to multi-gluon
amplitudes, and we review the Berends--Giele recursive relations,
both in their color-ordered and color-dressed form.
Section~\ref{sec:impl} describes our implementation of the various
techniques in practical algorithms, bound by the constraints of
existing computing platforms. The discussion of these
algorithms is continued in Sec.~\ref{sec:algos}, where we compare the efficiency
of sampling versus summing over color and helicity, and consider
computational and memory complexity.
We present a comprehensive analysis
of our timing studies in Sec.~\ref{sec:results} and draw conclusions
for future developments of event generators in Sec.~\ref{sec:outlook}.

\section{Efficient computation of multi-gluon tree amplitudes}
\label{sec:theory}
In this section, we recall various techniques for the efficient computation
of squared $n$-gluon matrix elements. They have a rich structure and are
an excellent proxy for more generic tree-level computations
typically encountered in collider physics.
The preferred technique for evaluating matrix elements
will depend on whether it is used for analytic calculations or numerical evaluation.
Furthermore, the optimal numerical evaluation technique depends
on the specific computational hardware.

\subsection{Summation over unobserved quantum numbers}
\label{sec:color_helicity_amplitudes}
Due to QCD confinement, it is impossible to detect spin or color of
individual partons. The effect of spin and color correlations
at the parton-level only becomes measurable through intricate
correlations in many-particle hadronic final-states.
Multi-parton amplitudes summed over color and spin are
therefore of highest interest for collider phenomenology.
At the same time, the summation over unobserved quantum numbers is one
of the major obstacles in the computation of collider physics
observables. We will first discuss how these computations are performed
efficiently.

We define the color and implicitly spin summed squared $n$-gluon tree-level amplitude
$\mathcal{A}_{a_1\ldots a_n}^{\mu_1\cdots\mu_n}$ as
\begin{equation}\label{eq:full_amplitude_decomposition}
  |\mathcal{A}(1,\ldots,n)|^2=
  \sum_{a_1\ldots a_n}
  \mathcal{A}_{a_1\ldots a_n}^{\mu_1\cdots\mu_n}(p_1,\ldots,p_n)
  \left(\mathcal{A}_{a_1\ldots a_n}^{\nu_1\cdots\nu_n}(p_1,\ldots,p_n)\right)^\dagger
  \prod_{i=1}^n P_{\mu_i\nu_i}(p_i,k_i)\;,
\end{equation}
where each gluon with label $i$ is characterized by its color $a_i$
and its momentum $p_i$. The projection operator $P_{\mu\nu}$ removes
the longitudinal component of the external gluons so no external ghosts
particles need to be included in the calculation. 
In the lightcone gauge the projection operator is given by
\begin{equation}\label{eq:tensor_axial}
  P_{\mu\nu}(p,k)=-g_{\mu\nu}+\frac{p_{\mu} k_{\nu}+p_{\nu} k_{\mu}}{p\cdot k}\;,
\end{equation}
where $k$ is an arbitrary massless gauge vector.
Motivated by the appearance of high transverse momentum production of hadrons
at hadron colliders and the discovery of 3-jet production 
at PETRA~\cite{Brandelik:1979bd,Berger:1979cj,Barber:1979yr},
the first analytic expressions for 4-gluon~\cite{Combridge:1977dm,Cutler:1977qm} 
and 5-gluon~\cite{Gottschalk:1979wq,Kunszt:1979ci} squared matrix elements 
were obtained by evaluating Eq.~\eqref{eq:full_amplitude_decomposition}
in terms of Feynman diagrams.

To simplify the calculation and to obtain compact expressions,
one can use the dyadic decomposition of the lightcone projection operator 
$P_{\mu\nu}$ into helicity eigenstates
\begin{equation}\label{eq:helicity_axial}
  P_{\mu\nu}(p,k)=\sum_{\lambda}\epsilon_{\mu}^{\lambda}(p,k)\epsilon_{\nu}^{\lambda\,\dagger}(p,k)\;,
\end{equation}
where $\lambda=\pm$ are the helicity labels of the gluon.
This was first used in~\cite{Berends:1981rb} to obtain a compact expression 
for the squared 5-gluon matrix elements by judicious choices 
of the gauge vectors $k_i$ to simplify the calculation. 
The squared $n$-gluon amplitude then takes the form
\begin{equation}
  \label{eq:full_amplitude_decomposition_discrete_summation}
 |\mathcal{A}(1,\ldots,n)|^2=\sum_{\lambda_1\ldots\lambda_n}\sum_{a_1\ldots a_n}
  \mathcal{A}_{a_1\ldots a_n}^{\lambda_1\ldots\lambda_n}(p_1,\ldots,p_n)
  \left(\mathcal{A}_{a_1\ldots a_n}^{\lambda_1\ldots\lambda_n}(p_1,\ldots,p_n)\right)^\dagger\;,
\end{equation}
where the $\mathcal A^\lambda$ are
the contraction of the original $\mathcal A^\mu$
with the helicity eigenstate $\epsilon_\mu^\lambda$.

The findings of multiple jet production at CERN 
(see e.g.~\cite{Physics3WorkingGroup:1987ndt,UA2:1991apcw})
motivated the calculation of 
6-parton helicity amplitudes using supersymmetry relations and
spinor algebra~\cite{Parke:1986gb,Gunion:1986zh,Gunion:1986cb}.
The obtained expressions were suitable for numerical evaluations,
but making them compact required a refined treatment of color.
The color decomposition introduced in~\cite{Berends:1987cv,Mangano:1987xk}
led to the definition of color-ordered amplitudes and
paved the way for more refined methods to compute helicity
amplitudes~\cite{Dixon:1996wi,Dixon:2013uaa}.

To evaluate Eq.~\eqref{eq:full_amplitude_decomposition_discrete_summation},
we have two choices
for the treatment of the helicity and/or color sums.
The traditional method
is to explicitly sum over all the quantum states.
The alternative technique is to replace the discrete
summation by a continuous sampling through parametric integrations.
For the spin states of the gluon this is familiar and involves changing helicities
to polarizations. The connection between the two approaches is 
straightforward~\cite{Draggiotis:1998gr,Draggiotis:2002hm}:
\begin{equation}\label{eq:polarization_kleiss}
  e_\mu(p,k,\phi)=e^{i\phi}\epsilon_{\mu}^+(p,k)+e^{-i\phi}\epsilon_{\mu}^-(p,k)\;,
\end{equation}
so that
\begin{equation}\label{eq:polarization_axial}
  P_{\mu\nu}(p,k)=\frac{1}{2\pi}\int_0^{\,2\pi}{\rm d}\phi\,
  e_\mu(p,k,\phi)e_{\nu}^{*}(p,k,\phi)\;.
\end{equation}
Instead of the complex polarization vector definition 
of Eq.~\eqref{eq:polarization_kleiss} one can also define a real valued polarization
vector (leading to real valued amplitudes for the gluon-only case).
We can choose a unit vector $e_1={\rm Re}(\epsilon^+)/2$
orthogonal to the gluon momentum $p$ and a subsequent unit vector $e_2=-{\rm Im}(\epsilon^+)/2$
orthogonal to $p$ and $e_1$. Then the continuous polarization vector is given by
\begin{equation}\label{eq:polarization_simple}
  e^\mu(p,\phi)=\cos(\phi)\,e_1^{\mu}(p)+\sin(\phi)\,e_2^{\mu}(p)\;.
\end{equation}

In a similar fashion we can replace the discrete sum over external colors 
by a continuous color polarization vector which is integrated over.
We first construct continuous color polarizations for the fundamental
representation based on the dyadic 
decomposition~\cite{Draggiotis:1998gr,Draggiotis:2002hm}
\begin{equation}
  \delta_{ij}=\int {\rm d}[z]\, \eta_{i}([z])\eta_{j}([z])\;.
\end{equation}
Any real valued $\eta_i$ can be parametrized in terms of polar
and azimuthal angles, such that
\begin{equation}
  \label{eq:color_angles}
  \eta_i([z])\to\eta_i(\theta,\phi)=\left(
  \begin{array}{c}
    \cos\theta \\
    \sin\theta\cos\phi \\
    \sin\theta\sin\phi \\
  \end{array}
  \right)\;,
\end{equation}
and
\begin{equation}
  \int {\rm d}[z]\to\frac{N_c}{4\pi}\int_0^{\,\pi}{\rm d}\cos\theta\int_0^{\,2\pi}{\rm d}\phi\;.
\end{equation}
In this way the quark color state is represented by a spherical three dimensional unit vector.

%One particularly convenient choice is to work with discrete indices.
%Formally, this corresponds to replacing the integral by~\cite{Maltoni:2002mq}
%\begin{equation}\label{eq:color_flow_def}
%  \delta_{ab}\to\sum_{i=1}^{N_c} \eta_{a}(i)\eta^\dagger_{b}(i)\;,
%  \qquad\text{where}\qquad
%  \eta_a(i)=\delta_{ia}\;.
%\end{equation}

In order to construct a dyadic decomposition for the adjoint representation,
relevant for gluons, we use the identity
\begin{equation}
  \begin{split}
    \delta^{ab}={\rm Tr}(T^aT^b)
    =&\;\int {\rm d}[z] {\rm d}[\bar{z}]\,\eta_{i}([z])T^a_{ij}\eta_{j}([\bar{z}])\, \eta_{k}([\bar{z}])T^b_{kl}\eta_{l}([z])\\
    =&\;\int {\rm d}[z] {\rm d}[\bar{z}]\,\eta^a([z],[\bar{z}])\eta^{b\,\dagger}([z],[\bar{z}])\;,
  \end{split}
\end{equation}
where we have defined the gluon color polarization vector
\begin{equation}\label{eq:gluon_color_polarization}
  \eta^a([z],[\bar{z}])=\eta_{i}([z])\,T^a_{ij}\,\eta_{j}([\bar{z}])\;.
\end{equation}
Note that this construction differs from~\cite{Draggiotis:1998gr,Draggiotis:2002hm},
in that it uses only four instead of five integration variables and the gluon
color polarization is represented by two real 3-dimensional spherical unit vectors.

Using the dyadic decompositions above, we can write the summed squared
matrix element in the following form:
\begin{equation}
  \left|\mathcal{A}(1,\ldots,n)\right|^2
  =\prod_i\sum\hspace*{-1.375em}\int\, \frac{{\rm d}\phi_i}{2\pi}\,
  {\rm d}[z]_i\,{\rm d}[\bar{z}]_i\;
  \left|\mathcal{A}_{[z]_1[\bar{z}]_1\ldots [z]_n[\bar{z}]_n}^{
    \phi_1\ldots \phi_n}(p_1,\ldots,p_n)\right|^2\;,
\end{equation}
where we have defined the color-helicity sub-amplitudes
\begin{equation}
  \begin{split}
  \mathcal{A}_{[z]_1[\bar{z}]_1\ldots [z]_n[\bar{z}]_n}^{
    \phi_1\ldots \phi_n}(p_1,\ldots,p_n)=
\mathcal{A}_{a_1\ldots a_n}^{\mu_1\cdots\mu_n}(p_1,\ldots,p_n)
\times\prod_ie_{\mu_i}(p_i,k_i,\phi_i)
\times\prod_j\eta_{a_i}([z]_i)\; .
\end{split}
\end{equation}
They can be further simplified analytically, or evaluated
numerically with the help of the Berends--Giele recursion.
Furthermore, the generation of continuous (color)
polarizations can easily be included in the Monte-Carlo
integration over phase-space. We will investigate these
questions in more detail in Secs.~\ref{sec:algos} and~\ref{sec:results}.

\subsection{Color decomposition and color-ordered amplitudes}
\label{sec:decomposition}
The decomposition of summed $n$-gluon amplitudes into helicity
amplitudes $\mathcal{A}^{\lambda_1\ldots\lambda_n}$ introduced
in Eq.~\eqref{eq:full_amplitude_decomposition}
is particularly useful in analytic calculations, because
it allows for a judicious choice of gauge vectors, which in turn
can greatly simplify the calculation~\cite{Berends:1981rb}.
Similarly, color decompositions are useful, because they allow
to factorize the color and kinematics dependent part of $n$-gluon
amplitudes and calculate the color-dependent coefficients once
and for all. This can be beneficial for both analytical and numerical 
evaluation. In this subsection, we will discuss color decompositions
that are particularly useful for fast numerical computation in
Monte-Carlo programs.

The most intuitive color decomposition of an $n$-gluon amplitude
$\mathcal{A}_n$ is based on the adjoint representation
of ${\rm SU}(3)$~\cite{Berends:1987cv,DelDuca:1999rs,DelDuca:1999ha}
resulting in
\begin{equation}\label{eq:color_decomposition_adjoint}
  \mathcal{A}_{a_1\ldots a_n}^{\lambda_1\ldots\lambda_n}(p_1,\ldots,p_n)
  =\sum\limits_{\vec\sigma\in S_{n-2}}
    (F^{a_{\sigma_2}}\ldots F^{a_{\sigma_{n-1}}})_{a_1a_n}\;
    A^{\lambda_1\ldots\lambda_n}(p_1,p_{\sigma_2},\ldots,p_{\sigma_{n-1}},p_n)\;,
\end{equation}
where $F^a_{bc}=if^{abc}$.
The functions $A$ are called color-ordered or partial amplitudes.
If they carry a helicity label they are often simply referred to
as helicity amplitudes.
The multi-index $\vec\sigma$ runs over all permutations $S_{n-2}$
of the ($n-2$) gluon indices $2\ldots n-1$.

Performing an explicit sum over colors in the squared amplitude,
Eq.~\eqref{eq:full_amplitude_decomposition}, leads to
\begin{equation}\label{eq:color_summed_analytic}
  \begin{split}
    |\mathcal{A}(1,\ldots,n)|^2=&\;\sum_{\lambda_1\ldots\lambda_n}\sum_{a_1\ldots a_n}
  \mathcal{A}_{a_1\ldots a_n}^{\lambda_1\ldots\lambda_n}(p_1,\ldots,p_n)
  \mathcal{A}_{a_1\ldots a_n}^{\lambda_1\ldots\lambda_n}(p_1,\ldots,p_n)^\dagger\\
  =&\;\sum_{\vec{\sigma},\vec{\sigma}'\in S_{n-2}}\mathcal{C}_{\vec{\sigma}\vec{\sigma}'}
  \sum_{\lambda_1\ldots\lambda_n}
  A^{\lambda_1\lambda_{\sigma_2}\ldots\lambda_{\sigma_{n-1}}\lambda_n}
  (p_1,p_{\sigma_2},\ldots,p_{\sigma_{n-1}},p_n)\\
  &\;\qquad\times A^{\lambda_1\lambda_{\sigma'_2}\ldots\lambda_{\sigma'_{n-1}}\lambda_n}
  (p_1,p_{\sigma'_2},\ldots,p_{\sigma'_{n-1}},p_n)^\dagger\;,
  \end{split}
\end{equation}
where the $(n-2)!^{\,2}$ color coefficients, $\mathcal{C}_{\vec{\sigma}\vec{\sigma}'}$,
are given by
\begin{equation}\label{eq:color_coefficient_adjoint}
  \mathcal{C}_{\vec{\sigma}\vec{\sigma}'}=\sum_{a_1\ldots a_n}
  (F^{a_{\sigma_2}}\ldots F^{a_{\sigma_{n-1}}})_{a_1a_n}
  (F^{a_{\sigma'_2}}\ldots F^{a_{\sigma'_{n-1}}})^{*}_{a_1a_n}\;.
\end{equation}
An alternative color decomposition can be obtained by using the
definition of structure constants in terms of ${\rm SU}(3)$
generators~\cite{Mangano:1987xk} 
\begin{equation}\label{eq:color_decomposition_fundamental}
  \mathcal{A}_{a_1\ldots a_n}^{\lambda_1\ldots\lambda_n}(p_1,\ldots,p_n)
  =\sum\limits_{\vec\sigma\in S_{n-1}}
    {\rm Tr}(T^{a_1}T^{a_{\sigma_2}}\ldots T^{a_{\sigma_n}})\;
    A^{\lambda_1\ldots\lambda_n}(p_1,p_{\sigma_2},\ldots,p_{\sigma_n})\;.
\end{equation}
In this case the sum runs over all permutations $S_{n-1}$
of the ($n-1$) gluon indices $2\ldots n$, leading to a substantial
increase in the number of partial amplitudes that contribute
to the full color-helicity amplitude. Equation~\eqref{eq:color_summed_analytic}
for the explicit sum is correspondingly changed by replacing
$S_{n-2}\to S_{n-1}$, $\lambda_n\to\lambda_{\sigma_n}$, $p_n\to p_{\sigma_n}$
and by making the corresponding replacements in the conjugate amplitude and
the color factors, Eq.~\eqref{eq:color_coefficient_adjoint}.
Note that the minimal number of partial amplitudes needed for a complete evaluation
is still $(n-2)!$~\cite{Kleiss:1988ne}. 
The remaining partial amplitudes can be obtained
from the calculated partial amplitudes by a set of linear equations.
The larger growth in the number of color coefficients
and the need to construct the remaining $(n-1)$ partial amplitudes
disfavors Eq.~\eqref{eq:color_decomposition_fundamental}
for numerical computations and we will therefore only use 
Eq.~\eqref{eq:color_decomposition_adjoint}.

A third color decomposition, suited especially for Monte-Carlo 
event generation, is the color-flow decomposition~\cite{Maltoni:2002mq}.
In this case, the gluon vector field is treated as an $N_c\times N_c$ matrix,
$A^\mu_{ij}$, rather than a field with one color index, $A^\mu_a$.
This leads to the amplitudes
\begin{equation}\label{eq:color_decomposition_colorflow}
  \begin{split}
  \mathcal{A}_{i_1j_1\ldots i_nj_n}^{\lambda_1\ldots\lambda_n}(p_1,\ldots,p_n)
  =&\;\prod_{k=1}^n T^{\,a_k}_{i_kj_k}
  \mathcal{A}_{a_1\ldots a_n}^{\lambda_1\ldots\lambda_n}(p_1,\ldots,p_n)\\
  =&\;\sum\limits_{\vec\sigma\in S_{n-1}}
    \delta^{i_1\bar{\jmath}_{\sigma_2}}
    \delta^{i_{\sigma_2}\bar{\jmath}_{\sigma_3}}
    \ldots\delta^{i_{\sigma_n}\bar{\jmath}_1}\;
    A^{\lambda_1\ldots\lambda_n}(p_1,p_{\sigma_2},\ldots,p_{\sigma_n})\;.
  \end{split}
\end{equation}
The maximum number of partial amplitudes to be evaluated for a
generic color configuration, $i_1j_1\ldots i_nj_n$, can be
as large as in Eq.~\eqref{eq:color_decomposition_fundamental}.
However, it can be shown that -- when combined with color sampling --
the color-flow decomposition yields the lowest average number
of partial amplitudes to be evaluated per Monte-Carlo
event at high parton multiplicity~\cite{Maltoni:2002mq}.

\subsection{Recursive computation of (color-)helicity amplitudes}
The computation of the color-ordered amplitudes can be accomplished using
the Berends--Giele recursion~\cite{Berends:1987me}, which corresponds to
a dynamic programming technique that efficiently caches sets of Feynman
diagrams with at least one common propagator. This propagator is dubbed
the off-shell current and is computed recursively as
\begin{equation}\label{eq:cobg}
  \begin{split}
    &J_\mu(1,2,\ldots,n)=\frac{-ig_{\mu\nu}}{p_{1,n}^2}\left\{\;
      \vphantom{\sum_{k=j+1}^{n-1}}
      \sum_{k=1}^{n-1}V_3^{\nu\kappa\lambda}(p_{1,k},p_{k+1,n})
        J_\kappa(1,\ldots,k)J_\lambda(k+1,\ldots,n)\right.\\
      &\qquad\qquad+\left.\sum_{j=1}^{n-2}\sum_{k=j+1}^{n-1}
        V_4^{\nu\rho\kappa\lambda}J_\rho(1,\ldots,j)
        J_\kappa(j+1,\ldots,k)J_\lambda(k+1,\ldots,n)\right\}\;.
  \end{split}
\end{equation}
Here $p_i$ denote the momenta of the gluons, $p_{i,j}=p_i+\ldots+p_j$ and
$V_3^{\nu\kappa\lambda}$ and
$V_4^{\nu\rho\kappa\lambda}$ are
the color-ordered three- and four-gluon vertices defined by
\begin{equation}
  \begin{split}
    V_3^{\nu\kappa\lambda}(p,q)&=i\,\frac{g_s}{\sqrt{2}}
      \big(\,g^{\kappa\lambda}(p-q)^\nu+
        g^{\lambda\nu}(2q+p)^\kappa-g^{\nu\kappa}(2p+q)^\lambda\,\big)\;,\\
    V_4^{\nu\rho\kappa\lambda}&=i\,\frac{g_s^2}{2}
      \big(\,2g^{\nu\kappa}g^{\rho\lambda}-
        g^{\nu\rho}g^{\kappa\lambda}-g^{\nu\lambda}g^{\rho\kappa}\,\big)\;.
  \end{split}
\end{equation}
The external particle currents, $J_\mu(i)$, 
are given by the helicity or polarization vectors
of Sec.~\ref{sec:color_helicity_amplitudes}.
A complete color-ordered $n$-gluon amplitude $A(1,\ldots,n)$ is
obtained by putting the helicity/polarization dependent ($n-1$)-particle
off-shell current $J_\mu(1,\ldots,n-1)$ on-shell and contracting it
with the external polarization $J_\mu(n)$:
\begin{equation}\label{eq:cobg_full_amplitude}
  A(1,\ldots,n)=J_\mu(n)\,p_{1,n}^2\,J^\mu(1,\ldots,n-1)\;.
\end{equation}
The algorithmic complexity scales as $\mathcal O(n^4)$
and is dictated by the four-gluon vertex.
This can be further improved by decomposing the four-gluon vertex and
introducing an auxiliary antisymmetric tensor field with the
``propagator''~\cite{Draggiotis:1998gr}
\begin{equation}\label{eq:prop_pseudo}
  -iD_{\mu\nu}^{\,\kappa\lambda}=
    -i\big(g_\mu^\kappa g_\nu^\lambda-g_\mu^\lambda g_\nu^\kappa\big)\;.
\end{equation}
Then the recursive relations for the gluon and tensor fields read
\begin{multline}\label{eq:cobg_tensor}
  J_\mu(1,2,\ldots,n)=\frac{-ig_{\mu\nu}}{p_{1,n}^2}
    \sum_{k=1}^{n-1}\left\{\vphantom{\sum_{k=1}^{n-1}}\;
      V_3^{\nu\kappa\lambda}(p_{1,k},p_{k+1,n})
      J_\kappa(1,\ldots,k)J_\lambda(k+1,\ldots,n)\right.\\
    {}+V_T^{\nu\kappa\alpha\beta}
      J_\kappa(1,\ldots,k)J_{\alpha\beta}(k+1,\ldots,n)
    +\left.V_T^{\lambda\nu\alpha\beta}J_{\alpha\beta}(1,\ldots,k)
      J_\lambda(k+1,\ldots,n)\vphantom{\sum_{k=1}^{n-1}}\right\}\;,
\end{multline}
and
\begin{equation}\label{eq:cobg_tensor_current}
  J^{\alpha\beta}(1,2,\ldots,n)=
    -iD^{\,\alpha\beta}_{\gamma\delta}\,
    \sum_{k=1}^{n-1}V_T^{\gamma\delta\kappa\lambda}
      J_\kappa(1,\ldots,k)J_\lambda(k+1,\ldots,n)\;,
\end{equation}
where the tensor-gluon interaction is given by
\begin{equation}\label{eq:cobg_tensor_vertex}
  V_T^{\mu\nu\kappa\lambda}=\frac{i}{2}\frac{g_s}{\sqrt{2}}
    \big(g^{\mu\kappa}g^{\nu\lambda}-g^{\mu\lambda}g^{\nu\kappa}\big)\;.
\end{equation}
This will improve the complexity scaling of the algorithmic implementation to $\mathcal O(n^3)$. 

The above recursion can be modified to calculate amplitudes 
without the need for an explicit color decomposition by 
using color-dressed currents instead.
The most convenient form for implementing the color-dressed recursion and
the techniques outlined in Sec.~\ref{sec:color_helicity_amplitudes} 
is given by the color-flow basis~\cite{Duhr:2006iq}, 
which works for both the discrete color
sampling strategy of~\cite{Maltoni:2002mq} and the continuous
sampling of~\cite{Draggiotis:1998gr,Draggiotis:2002hm}. It is based
on the use of the identity $\delta^{ab}={\rm Tr}(T^aT^b)$ for the gluon
propagator. By writing ${\rm Tr}(T^aT^b)=T^a_{ij}T^b_{ji}$ and assigning
color indices $i$ and $j$ to the intermediate gluon, one can use the
generators~$T^a_{ij}$ to project the structure constants associated
with the elementary interaction vertices onto fundamental indices as
\begin{equation}
  F^a_{bc}T^a_{ij}T^b_{kl}T^c_{mn}
  =if^{abc}T^a_{ij}T^b_{kl}T^c_{mn}
  =\delta_{il}\delta_{kn}\delta_{mj}
  -\delta_{in}\delta_{kj}\delta_{ml}\;.
\end{equation}
Note that we have used the conventions of~\cite{Dixon:1996wi} for the
normalization of the ${\rm SU}(3)$ generators.
Formally, we define the color-dressed gluon and tensor pseudoparticle currents
$\mathcal{J}_{\mu\,ij}$ and $\mathcal{J}_{\alpha\beta\,ij}$ as
\begin{equation}
  \label{eq:cdcurrents}
  \begin{split}
    \mathcal{J}_{\mu\,ij}(1,\ldots,n)&=
    \sum\limits_{\vec\sigma\in S_n}\delta_{ij_{\sigma_1}}
    \delta_{i_{\sigma_1}j_{\sigma_2}}
    \ldots\delta_{i_{\sigma_n}j}\;
    J_\mu(\sigma_1,\ldots,\sigma_n)\;,\\
    \mathcal{J}_{\alpha\beta\,ij}(1,\ldots,n)&=
    \sum\limits_{\vec\sigma\in S_n}\delta_{ij_{\sigma_1}}
    \delta_{i_{\sigma_1}j_{\sigma_2}}
    \ldots\delta_{i_{\sigma_n}j}\;
    J_{\alpha\beta}(\sigma_1,\ldots,\sigma_n)\;.
  \end{split}
\end{equation}
Denoting by $\pi$ the set $(1,\ldots,n)$ of $n$ particle indices,
the following recursive relations for these currents are obtained:
\begin{equation}\label{eq:color_dressed_currents}
  \begin{split}
    \mathcal{J}_{\mu\,ij}(\pi)&=
      D_{\mu\,ij}^{\,\nu\,hg}(\pi)\left\{\;
      \sum_{\mathcal{P}_2(\pi)}
        \mathcal{V}_{\nu\,hg}^{\,
	  \kappa\,kl,\,\lambda\,mn}(\pi_1,\pi_2)\,
        \mathcal{J}_{\kappa\,kl}(\pi_1)
        \mathcal{J}_{\lambda\,mn}(\pi_2)\right.\\
      &\text{\hspace*{18ex}}+\left.\sum_{\mathcal{OP}_2(\pi)}
        \mathcal{V}_{\nu\, hg}^{\,\kappa\,kl,\,\alpha\beta\,mn}\,
        \mathcal{J}_{\kappa\,kl}(\pi_1)
        \mathcal{J}_{\alpha\beta\,mn}(\pi_2)\;\right\}\;,\\[1em]
  \mathcal{J}_{\alpha\beta\,ij}(\pi)&=
    D_{\alpha\beta\,ij}^{\,\gamma\delta\,hg}\,
    \sum_{\mathcal{P}_2(\pi)}
    \mathcal{V}_{\gamma\delta\,hg}^{\,\kappa\,kl,\,\lambda\,mn}\,
    \mathcal{J}_{\kappa\,kl}(\pi_1)\mathcal{J}_{\lambda\,mn}(\pi_2)\;.
  \end{split}
\end{equation}
Here we have defined the color-dressed gluon and tensor pseudoparticle
vertices
\begin{equation}
  \begin{split}
    \mathcal{V}_{\nu\,hg}^{\,\kappa\,kl,\,\lambda\,mn}(\pi_1,\pi_2)=
      \delta_{lg}\delta_{kn}\delta_{hm}\,
      V_{3\,\nu}^{\;\;\,\kappa\lambda}(\pi_1,\pi_2)+
      \delta_{hk}\delta_{ml}\delta_{ng}\,
      V_{3\,\nu}^{\;\;\,\lambda\kappa}(\pi_2,\pi_1)
  \end{split}
\end{equation}
and
\begin{equation}
  \begin{split}
    \mathcal{V}_{\gamma\delta\,hg}^{\,\kappa\,kl,\,\lambda\,mn}=
      \delta_{lg}\delta_{kn}\delta_{hm}\,
      V_{T\,\gamma\delta}^{\;\;\,\kappa\lambda}+
      \delta_{hk}\delta_{ml}\delta_{ng}\,
      V_{T\,\gamma\delta}^{\;\;\,\lambda\kappa}\;,
  \end{split}
\end{equation}
as well as the dressed propagators
\begin{equation}
  {D_{\mu}^\nu}_{ij}^{hg}=D_{\mu}^\nu\,\delta_{ih}\delta_{jg}\;,
  \qquad\text{and}\qquad
  {D_{\alpha\beta}^{\gamma\delta}}_{ij}^{hg}=
  {D_{\alpha\beta}^{\gamma\delta}}\,\delta_{ih}\delta_{jg}\;.
\end{equation}
The first sum in Eq.~\eqref{eq:color_dressed_currents} runs over all
partitions $\mathcal{P}_2(\pi)$ of the set $\pi$ into two subsets $\pi_{1,2}$
while the second sum runs over the set of ordered partitions
$\mathcal{OP}_2(\pi)$ into $\pi_{1,2}$.\footnote{Our implementation of
  the sum over partitions in Eq.~\eqref{eq:color_dressed_currents} uses
  the equivalence of Eq.~(3.18) and Eq.~(3.16) in Ref.~\cite{Duhr:2006iq}
  to reduce the computational complexity.}

\section{Practical implementation of the algorithms}
\label{sec:impl}
In this section we discuss different implementations of the methods
laid out in Sec.~\ref{sec:theory} in actual computer code.
In most cases, we provide a CPU and a GPU version, which are analyzed
in terms of computational complexity and memory requirements in
Sec.~\ref{sec:algos} and compared in terms of practical compute
performance in Sec.~\ref{sec:results}. We use \rambo~\cite{Kleiss:1985gy}
to generate the phase-space points for the different methods. Obtaining a purely real algorithm
reduces the memory requirements by a factor of two compared to a complex valued algorithm.
In memory bound algorithms, reducing the memory of each object allows the CPU
and GPU to fetch more from the cache in a single read, thus improving overall performance.
Therefore, it is ideal to obtain a purely real implementation.

\subsection{Leading color computation}
The first algorithm (Tess) was originally published in~\cite{Giele:2011}:
\begin{itemize}
\item
  Helicity amplitudes are computed using the color-ordered Berends-Giele
  recursion, Eqs.~\eqref{eq:cobg} and~\eqref{eq:cobg_full_amplitude}.
  To avoid the larger memory requirements Tess does not make 
  use of the decomposed four-gluon vertex in
  Eq.~\eqref{eq:cobg_tensor}.
  Helicities are evaluated by sampling real polarization vectors
  according to Eq.~\eqref{eq:polarization_simple}.
  As a consequence, \emph{all} numbers in the algorithm are real.
\item
  Parallelization is achieved not only by calculating many events concurrently,
  but also by making use of the intrinsically parallel structure of the recursion,
  such that up to $n-1$ threads cooperate in the calculation of a single event.
\item
  Global memory access is minimized by storing only event weights and gluon
  momenta. The shared memory per event contains all the currents and momenta
  needed to compute the color-ordered amplitudes and scales quadratically with
  the number of particles.
\item
  In contrast to the original publication~\cite{Giele:2011},
  the implementation has been modified to use double precision numbers,
  as all of the following algorithms do.
\end{itemize}
The second algorithm (\newalgolc) is similar to \tess,
but differs in the following details:
\begin{itemize}
  \item
    Only event-level parallelization is used, i.e.\ every threads computes the
    matrix element for an independent phase-space point. 
  \item
    All quantities except for the internal currents of partial amplitudes are
    stored in the global memory of the GPU.
  \item
    Helicities can be either sampled or summed.
\end{itemize}
Both of the above algorithms rely on the leading color approximation.
The number of colors $N_c$ is considered to be large, such that the color sum
in Eq.~\eqref{eq:color_summed_analytic} can be evaluated analytically.
It results in a simple prefactor:
\begin{equation}
|\mathcal{A}_n(1, \ldots, n)|^{2} = N_c^{n-2}\left(N_c^{2}-1\right)
\left(
  \sum_{\vec \sigma \in S_{n-1}}\left|
  A^{\lambda_1\ldots\lambda_n}(p_1,p_{\sigma_2},\ldots,p_{\sigma_n})
  \right|^{2}+\mathcal{O}\left(\frac{1}{N_c^{2}}\right)
\right).
\end{equation}

\subsection{Color summation and sampling with color-ordered amplitudes}

The third and fourth algorithm (\newalgocosummed\ and \newalgocosampled)
use the following approach:

\begin{itemize}
\item
  Helicity amplitudes are computed by utilizing
  the color-ordered Berends--Giele recursion,
  Eqs.~\eqref{eq:cobg} and~\eqref{eq:cobg_full_amplitude}.
  The two \newalgoco\ algorithms do not make use
  of the decomposed four-gluon vertex in
  Eq.~\eqref{eq:cobg_tensor}, to avoid the larger memory requirements.
  Helicities are summed using real polarization vectors
  according to Eq.~\eqref{eq:polarization_simple}.
  As a consequence, \emph{all} numbers in the algorithms are real.
\item
  In the \emph{color summed} variant (\newalgocosummed),
  the adjoint representation decomposition from
  Eq.~\eqref{eq:color_decomposition_adjoint} is used,
  which leads to the amplitude assembly formula,
  Eq.~\eqref{eq:color_summed_analytic}.
  In the \emph{color sampling} variant (\newalgocosampled),
  the color-flow decomposition of
  Eq.~\eqref{eq:color_decomposition_colorflow} is used
  and color indices are generated event by event
  using the algorithm outlined in~\cite{Gleisberg:2008fv}.
\item
  All quantities except for the internal currents of partial amplitudes
  are stored in the global memory of the GPU.
  For the \newalgocosummed\ variant,
  this includes the color matrix,
  Eq.~\eqref{eq:color_coefficient_adjoint}, which we compute once using
  Form~\cite{KUIPERS20131453} and then read in from storage
  to perform the matrix element evaluations.
\end{itemize}
\subsection{Color sampling with color-dressed amplitudes}
The fifth algorithm (\newalgocdsampled) uses the following approach:
\begin{itemize}
\item
  Helicity amplitudes are computed using the color-dressed Berends--Giele
  recursion, Eq.~\eqref{eq:color_dressed_currents}, and in particular making use
  of the continuous color-polarizations introduced in
  Eq.~\eqref{eq:gluon_color_polarization}.
  Helicities are sampled using Eq.~\eqref{eq:polarization_kleiss},
  which is equivalent to Eq.~\eqref{eq:polarization_simple}.
  As a consequence, \emph{all} numbers in the algorithm are real.
\item
  All quantities are stored in global memory.
\end{itemize}

\section{Computational complexity and memory requirements}
\label{sec:algos}
In this section, we discuss the characteristics
of the various algorithms introduced in Sec.~\ref{sec:impl}
in terms of computational complexity and memory requirements
for practical implementations of Monte-Carlo programs,
which can be used for parton-level event generation
for collider experiments on GPUs. We will focus on
four criteria in particular:
i)~level of parallelization,
ii)~summing vs.\ sampling colors and helicities,
iii)~the scaling behavior of explicit summation and
iv)~memory requirements.
This guides us in our selection of a set of candidate algorithms
that will be studied in more detail in Sec.~\ref{sec:results}.
Note that unless explicitly mentioned, the timings are measured for
an event, meaning that the time to generate a phase-space point and evaluating
cuts is included. As can be seen in Fig.~\ref{fig:tess_vs_ours}, this
is only relevant for the lower multiplicities and becomes negligible
for the higher ones.

\subsection{Level of parallelization}
\label{sec:par}
The GPU implementation of the Tess algorithm
is not only parallelized at the event level,
but also at the lower level of the summation
over currents in the recursive amplitude computation~\cite{Giele:2011}.
This corresponds to a parallel computation of the (outer) sums in Eq.~\eqref{eq:cobg},
such that up to $n-1$ threads can cooperate.
The low-level parallelism can be helpful in a memory bound algorithm
with several threads operating on the same data,
which can be stored on the shared memory of the GPU,
a fast cache managed by the program itself.
The downside is a more complex implementation,
due to the data sharing and the combination of different
levels of parallelization.

\begin{figure}[htbp]
  \centering
  \includegraphics[width=.7\textwidth]{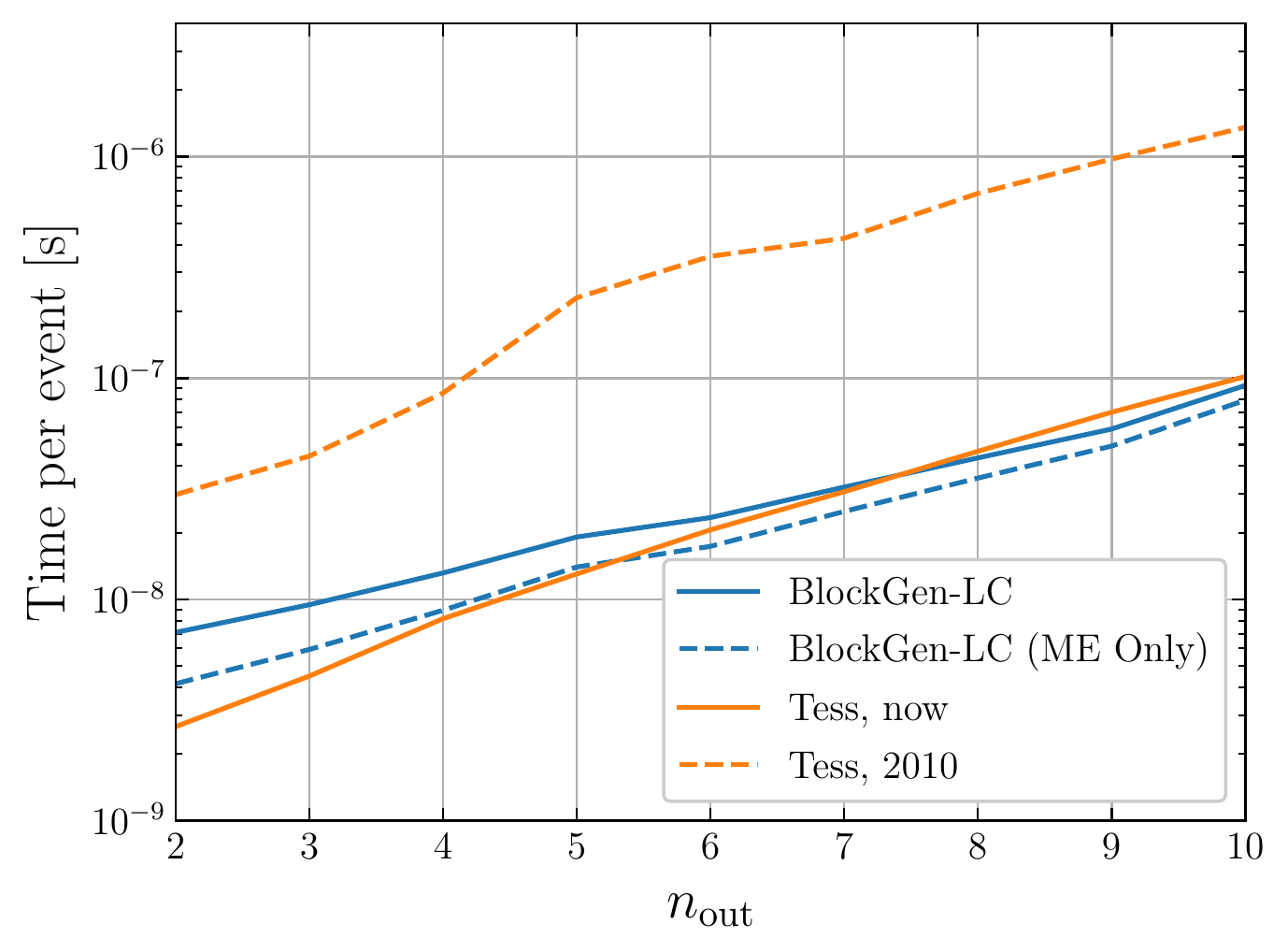}
  \caption{A comparison of the runtimes of \tess\ and \newalgolc,
    measured on a NVIDIA V100 (\SI{16}{GB} global memory, 5,120 CUDA cores, \SI{6144}{KB} L2 cache). To make a more fair comparison to \tess,
    we include the timing for \newalgolc\ only including the time to
    evaluate the matrix element (``ME Only''). The results for \tess\ from
    the original publication in 2010~\cite{Giele:2011} are included as
    a reference to show the GPU improvements.}
  \label{fig:tess_vs_ours}
\end{figure}

Modern HPC GPU combine shared memory with the chip-controlled L2 cache,
and have a larger throughput for loading missing data from their global memory.
Therefore, the speed gains from the additional level of parallelization
might be diminished.
We analyze this effect in Fig.~\ref{fig:tess_vs_ours} which shows the computational time
per event comparing Tess (orange) to \newalgolc\ (blue).
For high gluon multiplicities, the two implementations perform similarly.
The different scaling of \newalgolc\ at low multiplicities arises
from the slightly more generic handling of phase-space generation and cuts in \newalgolc\ and is removed
when we adopt the same strategy as Tess. This is shown in dashed blue.
For this reason, we only consider pure event-level parallelization
in all the following implementations, avoiding unnecessary complexity.
It is interesting to note the large timing improvement in Fig.~\ref{fig:tess_vs_ours}
when comparing the original results from Tess (dashed orange) to the current results (orange).
This improvement is purely due to improvements in GPU hardware, and indicates
the increasing importance of modern GPUs for collider physics applications.

\subsection{Efficiency of color and helicity sampling}
\begin{figure}[htbp]
  \centering
  \includegraphics[width=.7\textwidth]{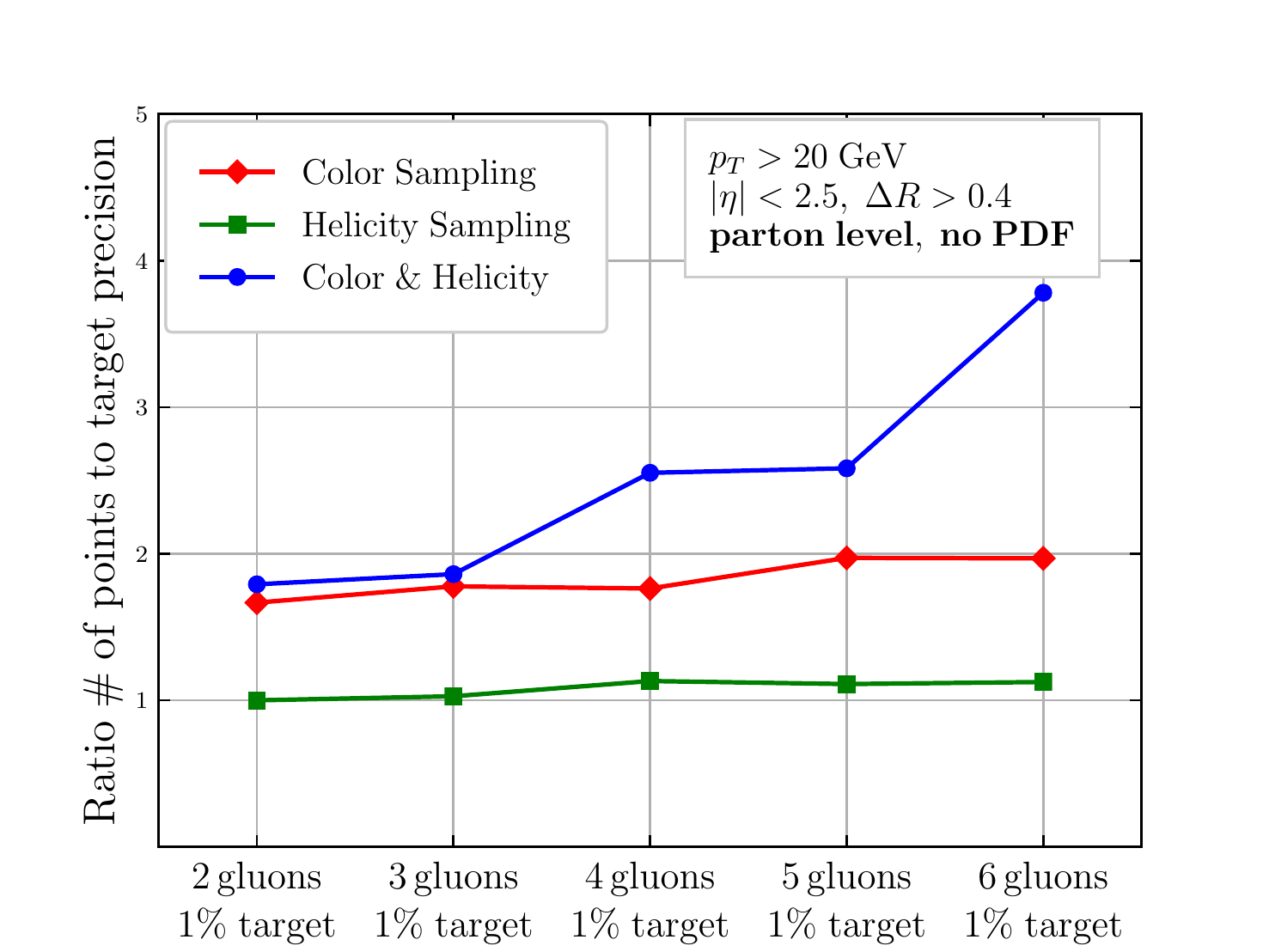}
  \caption{A comparison of the overhead for color, helicity, and combined color-helicity
    sampling as a function of gluon multiplicity.
    The ratio of the number of points needed
    to reach a target precision of 1\% is shown relative to a calculation
    with an explicit sum over both colors and helicities.}
  \label{fig:overhead_points}
\end{figure}
Figure~\ref{fig:overhead_points}
shows the overhead for color sampling, helicity sampling, and combined
color and helicity sampling as a function of gluon multiplicity
in a realistic parton-level calculation.
For this, we use the recursive phase-space generator of \comix.
Other details of the setup are the same as in Sec.~\ref{sec:results}.
We require a target precision of 1\%. We measure the required number
of points needed to attain this precision and plot the ratio of the
number of points for the various sampling methods to the explicit sum.
We observe that the relative overhead of helicity and color sampling
individually is approximately constant as a function of multiplicity.
However, the overhead of combined color and helicity sampling increases
with increasing multiplicity. This can be understood by looking at the
analytic structure of the $n$-gluon amplitudes. For $2$ and $3$ gluon
final states, only MHV amplitudes exist. The numerator of a
mostly plus amplitude is given as the fourth power of the product
of the spinors for the two gluons with negative helicity. Therefore,
helicity configurations can be sampled efficiently by drawing negative
helicity pairs with probability proportional to the magnitude of the
two-particle invariants. For higher multiplicites, the number of MHV
configurations is still larger than the number of NMHV configurations,
but the sampling according to two-particle invariants does not always
approximate the true structure of the matrix element. This effect
seems less pronounced in the color summed case, as can be seen in
Fig.~\ref{fig:overhead_points}. However, when colors are sampled,
an imperfect selection of the helicity configuration does create large
fluctuations in $4$- and $5$-gluon final states (where MHV and NMHV
configurations are present), and yet larger fluctuations in $6$-gluon
final states (where MHV, NMHV and NNMHV configurations exist).\footnote{
  We were not able to fully confirm the expected behavior in
  $7$-gluon final states, because the corresponding color-summed
  prediction could not be determined within our computing budget.
  However, the ratio between color-helicity summed and color summed
  only prediction is indeed the same as in the case of the
  $6$-gluon final state.}
The reason for this is that the numerator and denominator of the
partial amplitudes are uncorrelated, thus an uncorrelated and
imperfect sampling of color and helicity leads to reduced efficiency.
Nevertheless, the timing improvements in the matrix element computation
that arise from the use of sampling algorithms will never be
overcompensated by a reduced convergence in the eventual
Monte-Carlo integration.
This point is crucial when making a choice about which algorithm
to implement for GPU computing, and in particular it enables us
to choose a memory-lean option.

\subsection{Scaling of color and helicity sums}
\label{sec:summation_scaling}

When evaluating the sum over pairs of permutations in Eq.~\eqref{eq:color_summed_analytic}
(for a given helicity configuration),
the $(n-2)!$ matrix elements needed should first be calculated and stored.
Then the $(n-2)!((n-2)!+1)/2$
independent summands can be calculated
without re-evaluating the same matrix elements over and over.
The time needed to evaluate each summand is then dominated
by loading the data corresponding to $\mathcal C_{\vec\sigma\vec\sigma'}$,
$A^{\vec\sigma}$ and $A^{\vec\sigma'}$,
and should hence be individually very small compared
to the evaluation of a partial amplitude via the recursion relations in Eq.~\eqref{eq:cobg}.
However, the different scalings  of the two operations,
factorial for precalculating partial amplitudes vs.\ factorial squared for evaluating the color sum,
have the consequence that there will be an $n_\text{out}$
for which the time required for the summation will eventually become dominant.

\begin{figure}[htbp]
  \centering
  \includegraphics[width=0.7\textwidth]{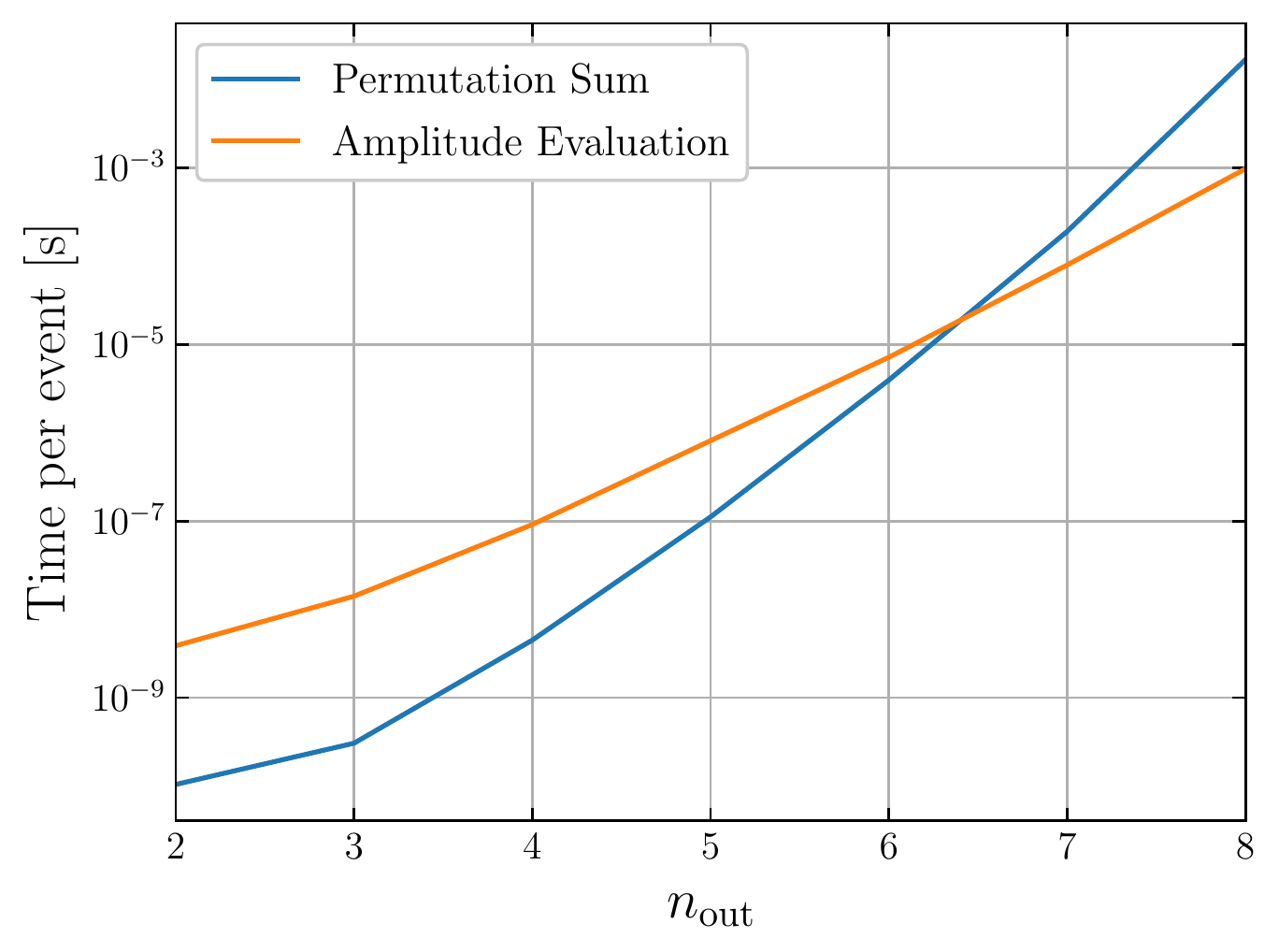}
  \caption{A comparison of the time needed to evaluate the required amplitudes
    versus the time needed to sum over all the permutations required as a function
    of multiplicity (Note that the absolute values are not directly comparable
    to Fig.~\ref{fig:single_threaded_timing}, since we have used helicity sampling here.)}
  \label{fig:scaling_me_vs_sum}
\end{figure}

Figure~\ref{fig:scaling_me_vs_sum}
compares the time needed per event for each operation.
The crossover point is found to be between $n_\text{out}=6$ and $7$.
The inherent $\mathcal O((n-2)!^{\,2})$ scaling of the color-summed algorithm
means that a color-sampled algorithm will eventually become more efficient.
However, since the contribution of the summing itself becomes relevant
only beyond $n_\text{out}=6$,
this behavior might not affect practically relevant computations.

\subsection{Memory requirements of summed vs.\ sampled algorithms}
\label{sec:mem}
\begin{figure}[htbp]
  \centering
  \includegraphics[width=0.7\textwidth]{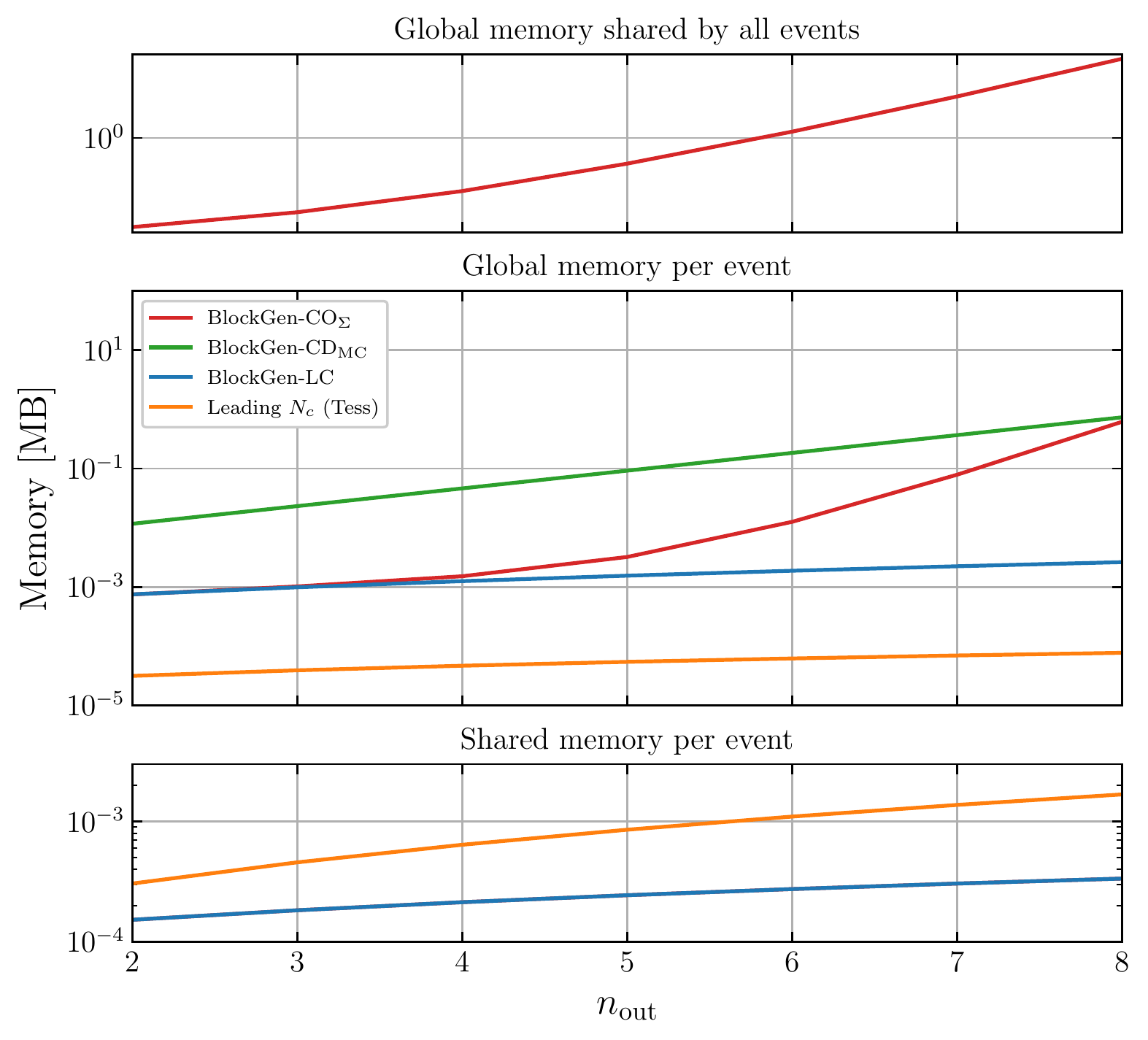}
  \caption{Comparison of memory usage in different GPU implementations.
    The upper plot displays the global
    memory required that is allocated independently of the number of events/threads.
    The middle plot shows the global memory per event
    and the bottom plot shows the shared memory per event. The shared memory per
    event for \newalgolc\ and \newalgocosummed\ is identical.}
    \vspace{2mm}
  \label{fig:compare_gpu_memory}
\end{figure}

Figure~\ref{fig:compare_gpu_memory} compares
the different memory requirements on the GPU for the three different
full-color algorithms \newalgocosummed, and \newalgocdsampled.
It also includes
leading color results from Tess and \newalgolc.
The upper plot displays the global memory usage which is not event-specific,
the middle plot shows the global memory usage per event, and
the bottom plot shows the shared memory usage per event.

\begin{figure}[htbp]
  \centering
  \includegraphics[width=0.7\textwidth]{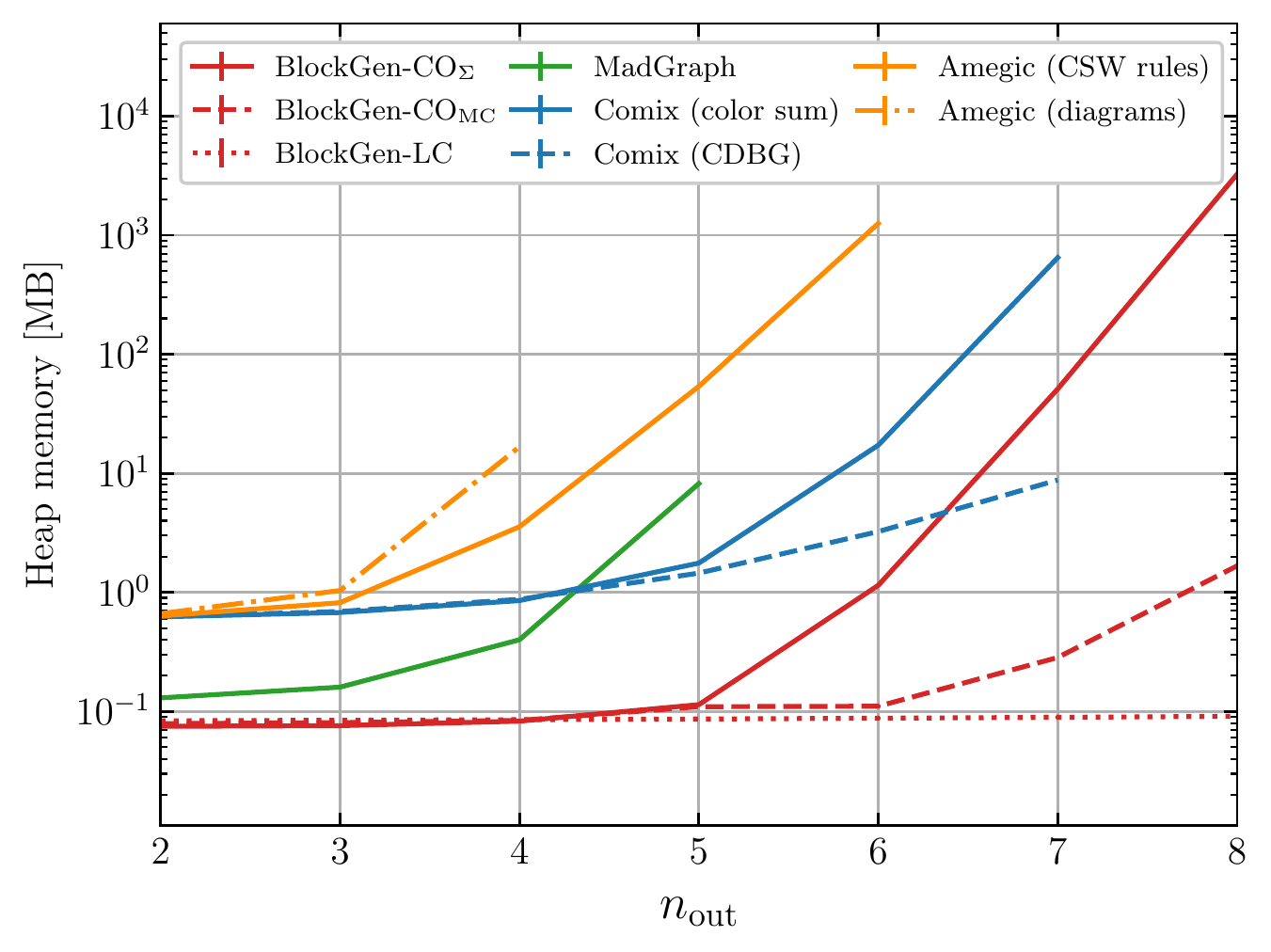}
  \caption{Single threaded CPU heap memory usage for the various algorithms (Note that the MadGraph 
    executable size is used, since the color matrix, etc.\ is compiled into
    the executable. Therefore, the MadGraph dynamic heap memory allocations 
    are negligible).
  }
  \label{fig:single_threaded_memory}
\end{figure}

The Tess algorithm uses global memory only to
store event weights and gluon momenta,
which gives a small footprint in the per-event global memory plot.
Everything else is stored in shared memory.

In the \newalgocosummed\ algorithm,
event-independent global memory is used
in particular to store the
color matrix, Eq.~\eqref{eq:color_coefficient_adjoint},
leading to a rapid $\mathcal O((n-2)!^{\,2})$ growth.
The per-event global memory usage
is dominated by the $\mathcal O((n-2)!)$ growth of the number of color-ordered
amplitudes $A(1,{\sigma_2},\ldots,{\sigma_{n-1}},n)$.
Shared memory is used only for the intrinsic momenta in the recursion,
such that its use grows linearly with the number of particles.

For \newalgocdsampled, the per-event global memory use is governed
by the number of complex valued currents and tensors. 
Since we only need to compute the currents for ordered sets of $\pi$
in Eq.~\eqref{eq:color_dressed_currents}, the total number of objects
scales as $\mathcal O(2^n)$.
Shared memory is not used by the color-dressed approach.

We first note that the color-summed algorithm
is limited to $n_\text{out} \leq 8$. Beyond that
the memory required to store the color matrix will
exceed $\mathcal O(\SI{10}{\giga\byte})$, and hence
will not fit any longer into the global memory of the GPU.\footnote{
  At the expense of relabeling permutations we could
  minimize the size of the storage required for the color matrix
  and obtain a scaling of $\mathcal O((n-2)!)$,
  down from $\mathcal O((n-2)!^{\,2})$.
  However, this technique would be beneficial only for large final-state
  multiplicities which are generally better to compute in a
  color-dressed approach due to the improved scaling.
  We therefore refrain from using it.
}
However, for $n_\text{out} \lesssim 7$ the memory use
of the color-summed algorithm is below that of the color-dressed algorithm.
Given that the algorithms are memory bound,
we expect the color-summed algorithm to perform better
than the color-dressed one for $n_\text{out} \lesssim 7$.

Figure~\ref{fig:single_threaded_memory} compares
the single-thread memory usage
of CPU-based implementations of our algorithms
\newalgocosummed, \newalgocosampled\ and \newalgolc,
with the widely used
\madgraph, \comix, and \amegic\ codes.
For \madgraph, we use version 2.9.2,
while \comix\ and \amegic\ have been run
as part of the \sherpa\ framework
in its version 2.3 series~\cite{Bothmann:2019yzt}.
For \madgraph, we use the standalone mode,
supplemented with a custom Monte-Carlo loop.
For \comix, we use its color summing and its color sampling mode,
while for \amegic, we show the default diagram-based mode,
and a mode where analytic CSW rules are employed~\cite{Gleisberg:2008ft},
which have been derived in~\cite{Cachazo:2004kj}.
We plot the RAM memory use of all codes except for \madgraph,
where the code generator embeds the color matrices directly
within the source code,
such that the memory usage is nearly exclusively driven
by the size of the compiled executable,
which we therefore plot instead in this case.
%Figure~\ref{fig:single_threaded_memory}
%also contains subplots for the ratios of all color summing codes to our
%\newalgocosummed\ and of all color sampling codes to our
%\newalgocosampled.
We find that the \comix\ modes and our custom algorithms
have the smallest memory footprints for multiplicities
of $n_\text{out}=5$ and beyond,
with \comix\ in its summed and its sampled mode using
about 10 and 25 times more memory than the respective \newalgoco.
This is likely due to \comix\ storing all contributing sub-currents
of Eq.~\eqref{eq:cdcurrents}, to reduce evaluation time~\cite{Gleisberg:2008fv}.

%Comparing the color-sampled code with \comix\ (CDBG), we find that \comix\
%requires more memory but features a better scaling behavior with respect to
%the number of outgoing particles. Some of the additional memory usage of
%\comix\ is due to the overhead that comes with the deployment in a more
%general setup in \sherpa. The remaining memory overhead is mostly caused by
%storing all the contributing sub-currents. The worse scaling of our
%color-sampled code is caused by storing all the possible permutations
%resulting in a factorial growth.

\section{Performance of event generation}
\label{sec:results}

We now compare the algorithms
described in the previous sections
in terms of their practical performance.\footnote{
  CUDA Nsight profiling reports can be obtained from the authors upon request.}
Of particular interest is the time needed to generate a single event.
Where appropriate, we include commonly used tools in the comparison,
such as \madgraph~\cite{Alwall:2011uj}, \amegic~\cite{Krauss:2001iv}
and \comix~\cite{Gleisberg:2008fv}
(the versions used are listed in Sec.~\ref{sec:mem}).

We begin by studying single-thread performance on CPU in Sec.~\ref{sec:results_cpu},
proceed with the massively parallelized calculation on GPU in Sec.~\ref{sec:results_gpu},
and conclude with a realistic chip-to-chip comparison
of (multi-threaded) CPU and GPU event generation in Sec.~\ref{sec:results_cpu_vs_gpu},
in order to find the most promising hardware/algorithm combination.

For all studies, we use a partonic center-of-mass energy of $\sqrt{s}=\SI{14}{\TeV}$
and require the gluon momenta to satisfy the kinematic constraints
\begin{equation*}
  p_{T}>20~{\rm GeV}\;,\qquad
  |\eta|<2.5\;,\qquad
  \Delta R>0.4\;.
\end{equation*}
A parton density function is not used.
For simplicity, we set the strong coupling to $\alpha_S(m_Z)=0.118$
and use a fixed renormalization scale of $\mu_R=m_Z$.\footnote{
  Even though we only study gluon amplitudes, none of our algorithms
  makes use of the process-specific symmetries relating different helicity
  amplitudes. We do so in order to reflect the performance that can be expected
  from a generic, automated implementation of an event generator for arbitrary
  interaction model Lagrangians.
}

\subsection{Comparison of single-threaded algorithms}
\label{sec:results_cpu}
\begin{figure}[t]
  \centering
  \includegraphics[width=0.8\textwidth]{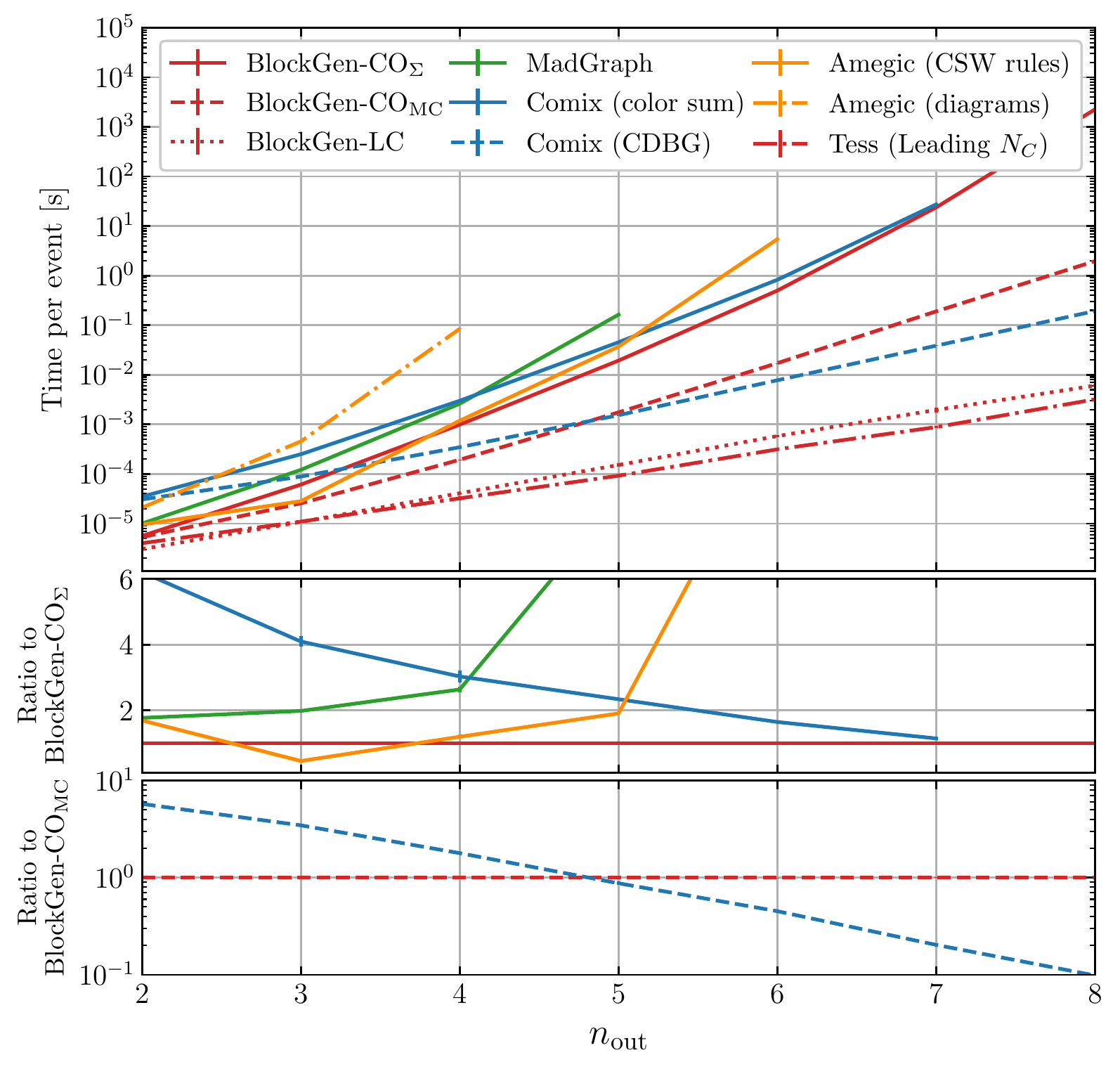}
  \caption{The timings for various CPU-based algorithms run on a single thread are compared
    as a function of the gluon multiplicity.
    The results were all generated on an Intel$^\text{\textregistered}$
Xeon$^\text{\textregistered}$ E5-2650 v2 8-core CPU (\SI{2.60}{GHz}, \SI{20}{MB} cache).}
  \label{fig:single_threaded_timing}
\end{figure}

To study the evaluation time per event of the algorithms
without parallelization,
we compare in Figure~\ref{fig:single_threaded_timing}
the CPU variants of our algorithms
\newalgocosummed, \newalgocosampled\ and \newalgolc,
with the widely used public codes
\madgraph, \comix, and \amegic\ ME generators,
and with \tess.
As described in Sec.~\ref{sec:mem},
\comix\ is used in color summing and sampling mode,
while \amegic\ is using either its default diagrammatic mode
or analytic CSW rules.
Figure~\ref{fig:single_threaded_timing} shows the evaluation time per
event for these different algorithmic choices, as well as for the public codes.
We find that the leading color implementations, \newalgolc\ and \tess, perform similarly.
For the full-color algorithms,
ratio plots show the evaluation times relative to \newalgocosummed\ and
\newalgocosampled. Note that the implementation of \newalgocosummed\ and
\newalgocosampled\ is based on purely real numbers, cf.\ Sec.~\ref{sec:impl}.
A certain overhead compared to the public codes, which work with complex numbers,
is therefore to be expected.

In both cases, \comix\ has the best scaling behavior due to the reuse of
a maximal number of precomputed sub-currents. This reduces the naively
expected factorial growth in computing time to an exponential growth, but
comes at the cost of a larger memory footprint (see Sec.~\ref{sec:mem}).
Note that the recycling of sub-currents is not a practical option for the
color-ordered algorithms. While the color-dressed algorithm only requires
to store one current per \textit{unordered set} of indices, the color-ordered
algorithms (both in the color summed and color sampled variant) would require
to store \textit{ordered sets} of indices. As the number of ordered sets
grows factorially with the number of members in the set, the memory required
for this technique would become prohibitively large at large multiplicity.
The resulting different scaling behaviors are clearly visible
in Fig.~\ref{fig:single_threaded_timing}.
Hence, \comix' evaluation times are smallest for large multiplicities,
falling below \newalgocosummed\ at ${n_\text{out} \approx 7}$
and below \newalgocosampled\ at $n_\text{out} \approx 5$,
for summing and sampling, respectively.

Taking Fig.~\ref{fig:overhead_points} as a ballpark reference for how much a color sampled
algorithm might be penalized in terms of accuracy with respect to a color summed one,
we can also conclude that for single-threaded algorithms, color sampling
is much faster than summing for large multiplicities.

\subsection{Comparisons of massively parallel execution on GPU}
\label{sec:results_gpu}
\begin{figure}[t]
  \centering
  \includegraphics[width=0.7\textwidth]{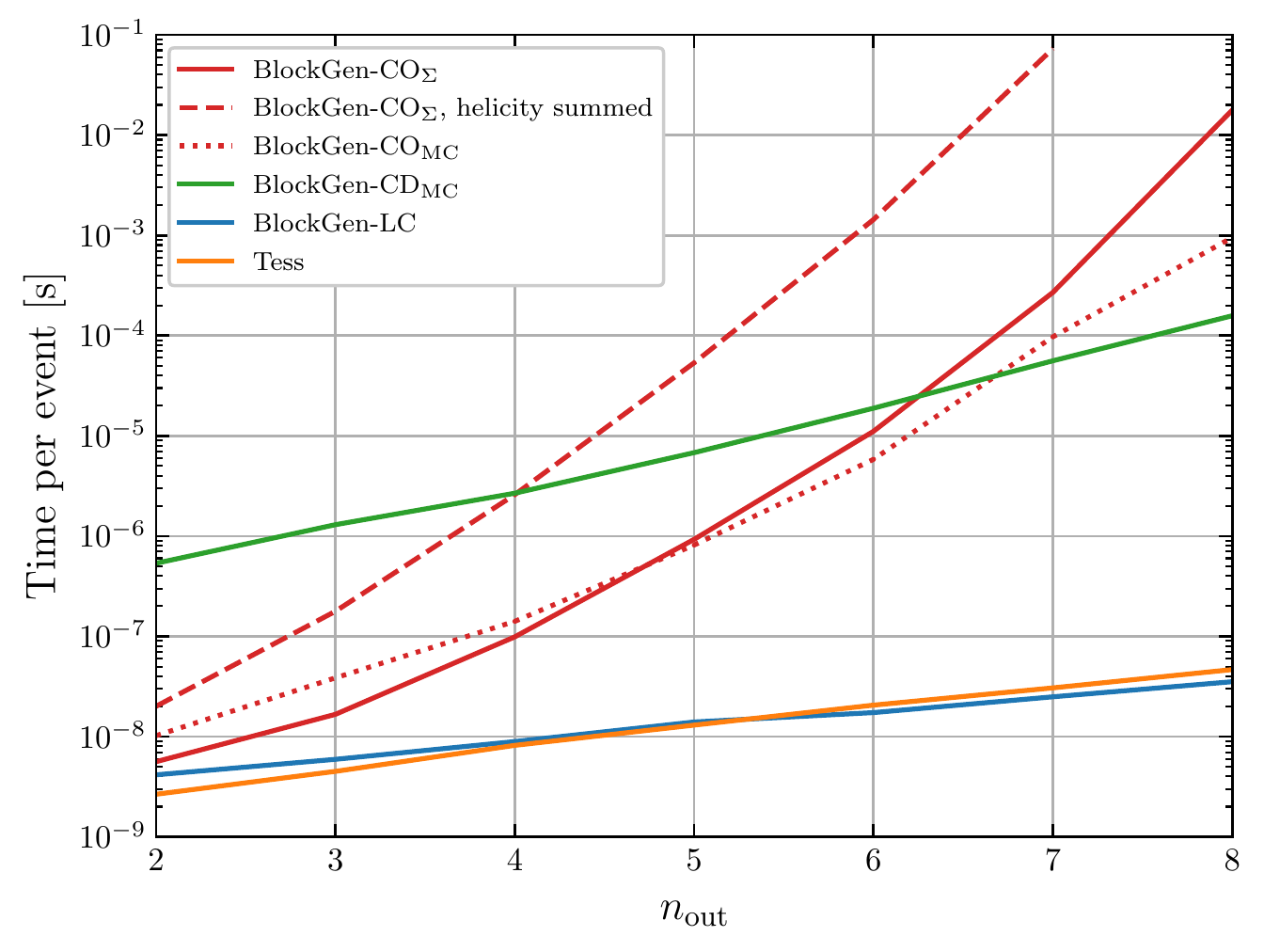}
  \caption{The timings for various GPU-based algorithms are compared as a function
    of gluon multiplicity. All algorithms were run on an NVIDIA V100 (\SI{16}{GB} global memory, 5,120 CUDA cores, \SI{6144}{KB} L2 cache).}
  \label{fig:compare_gpu_codes}
\end{figure}

Figure~\ref{fig:compare_gpu_codes} displays the evaluation time per event for
our GPU codes as introduced in Sec.~\ref{sec:impl}.

As opposed to the single thread results in Sec.~\ref{sec:results_cpu},
we now observe a similar performance for
the color-summed \newalgocosummed\ and the color-sampled \newalgocdsampled\ algorithms.
This is caused by the varying number of valid color configurations
for each thread in the color-sampled case:
Threads with few valid configurations need to wait for those with many.
However, the scaling of the color-sampled version still is better,
which is mainly due to the absence of the $\mathcal O((n-2)!^{\,2})$
scaling of the color summation
as discussed in Sec.~\ref{sec:summation_scaling}.

The color-dressed \newalgocdsampled\ exhibits an excellent scaling,
but is offset by almost two
orders of magnitude and only performs better than the color-ordered codes for
${n_{\mathrm{out}} \gtrsim 6}$. As can be seen in Eq.~\eqref{eq:cdcurrents},
the color-dressed approach requires significantly more memory per current,
as discussed in Sec.~\ref{sec:mem},
such that the memory handling requires a significant amount of time during execution.

%where in particular the thread independent memory is an order of magnitude
%larger in the color dressed approach than in the color summed version. As
%discussed in Sec.~\ref{} we primarily use global memory on the GPU's which
%penalizes this memory heavy algorithm.
%

\subsection{Comparison of single-threaded and massively parallelized algorithms}
\label{sec:results_cpu_vs_gpu}

To compare the CPU-based with the massively parallelized GPU-based algorithms,
we run the single-threaded \amegic\ (with CSW rules enabled) and \comix\ algorithms
parallelized via MPI (Message Passing Interface) over 16 threads.
The resulting performance is very close to dividing the single-thread evaluation times by 16.
By making use of all the threads of our test CPU,
we achieve a realistic chip-to-chip comparison to the performance of our GPU-based algorithms.
However, we note that the CPU is a discontinued Intel$^\text{\textregistered}$
Xeon$^\text{\textregistered}$ E5-2650 v2 (\SI{2.60}{GHz}, \SI{20}{MB} cache), while the GPU is a
modern NVIDIA V100 (\SI{16}{GB} global memory, 5,120 CUDA cores, \SI{6144}{KB} L2 cache)\footnote{For additional information about the NVIDIA V100, see \url{https://images.nvidia.com/content/volta-architecture/pdf/volta-architecture-whitepaper.pdf}}. In order to emulate a realistic chip-to-chip comparison for a similarly
capable hardware, we would need to scale the number of physical CPU cores from 8 to 36, corresponding
for example to an Intel$^\text{\textregistered}$ Xeon$^\text{\textregistered}$ Platinum 8360Y
(\SI{2.40}{GHz}, \SI{54}{MB} Cache).
Moreover, we note again that the implementation
of \newalgocosummed\ and \newalgocosampled\ is based
on purely real numbers, cf.\ Sec.~\ref{sec:impl}.
We therefore expect an overhead of about a factor of two
compared to the public codes, which work with complex numbers.

\begin{figure}[t]
  \centering
  \includegraphics[width=0.8\textwidth]{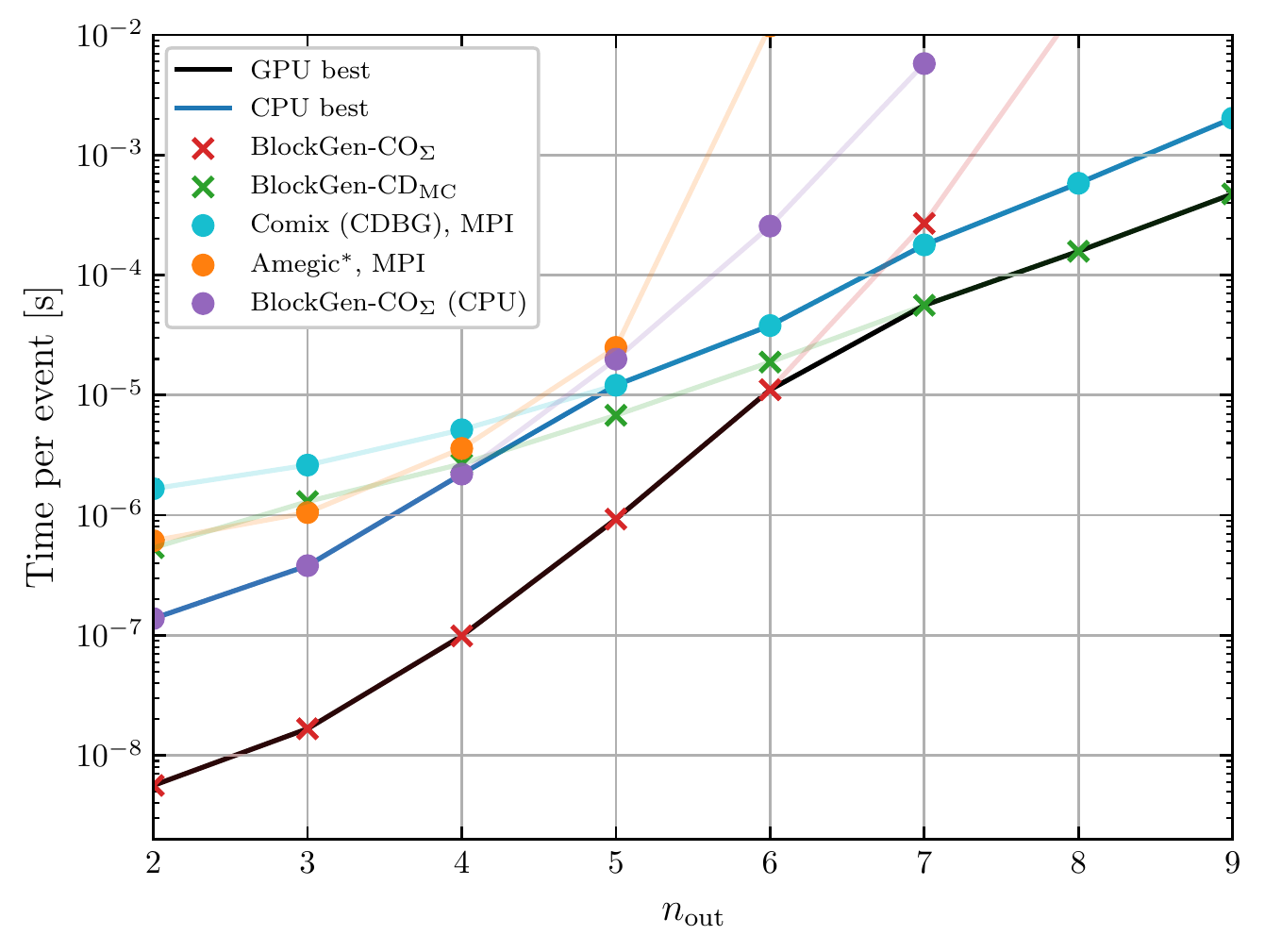}
  \caption{The timings for GPU-based (crosses) and CPU-based (dots) algorithms are compared against each other
    as a function of gluon multiplicity. The CPU numbers are all generated on
    an Intel$^\text{\textregistered}$ Xeon$^\text{\textregistered}$ E5-2650 v2 8-core
    CPU, (\SI{2.60}{GHz}, \SI{20}{MB} cache), while all the GPU numbers are generated on a NVIDIA V100 (\SI{16}{GB} global memory, 5,120 CUDA cores, \SI{6144}{KB} L2 cache). The MPI
    versions are run on 16 threads, and the timing for the color summed algorithm is divided by
    a factor of 16 to mimic the improvements that would occur from MPI.
    Furthermore, a modified version of Amegic is used in order to perform helicity sampling.}
  \label{fig:cpu_vs_gpu}
\end{figure}

The resulting times are displayed in
Fig.~\ref{fig:cpu_vs_gpu}. We find that for
$n_{\mathrm{out}} \leq 6$, the best-performing GPU algorithm (\newalgocosummed) has
evaluation times per event which are at least an order of magnitude smaller than
the best CPU evaluation times. Taking into account hardware differences, the
improvement is reduced by about a factor of four, but remains substantial.
For $n_{\mathrm{out}} = 7$, the GPU code performs similarly to the CPU code.
For $n_\text{out} \geq 7$,  \newalgocdsampled\ becomes the fastest algorithm.
However, \comix\ achieves a similar performance on the CPU in that region,
when accounting for hardware differences and the usage of complex instead of real numbers.

\section{Conclusion}
\label{sec:outlook}
As a first step towards GPU-assisted Standard Model Monte-Carlo event generators,
this study explored a variety of methods to calculate leading order $n$-gluon squared
amplitudes with full color dependence. In view of increased precision requirements
for high-multiplicity final-state observables at the high-luminosity LHC, we focused
in particular on the scaling of the computation time with the number of external gluons.
We studied different algorithms based on the Berends--Giele recursion, which differ in the
treatment of helicity and color sums involved when squaring the scattering amplitude.
The study was performed within a set of constraints relevant for realistic simulations.
In particular, we did not take advantage of special symmetries in all-gluon scattering
amplitudes and only considered algorithms that generate strictly positive weights.
Our results are therefore representative of what can be expected from a full-fledged
matrix-element calculator in the Standard Model and extensions thereof, which do not
involve more than four-particle interactions in the Lagrangian.

Since the different algorithms vary in their scaling and memory access patterns, the
algorithm with optimal performance depends on the final-state particle multiplicity.
Therefore, there is no one best choice for LHC phenomenology.
At low to medium multiplicity,
explicit color summation combined with color ordered amplitudes, as implemented in
\newalgocosummed, provides the best performance. At high multiplicity, color sampling
with color dressed amplitudes, as implemented in \newalgocdsampled, is preferred
due to the different scaling pattern. It is interesting to note that in a direct chip-to-chip
comparison the existing parallel CPU-based \comix\ generator
performs at a similar level as \newalgocdsampled\ at high multiplicity.
Given that typical event simulations
at the LHC will not require more than seven jets in the final state, we conclude
that \newalgocosummed\ is the optimal choice to implement as a full-fledged low-to-medium
multiplicity event generator on the GPU.

The next steps towards a complete Monte Carlo simulation of Standard Model events
will involve the addition of quark processes and the development of a GPU-based
phase-space generator. Since high-performance computing systems have both CPUs and GPUs
associated with each node, it will be interesting to explore hybrid computation
schemes with load balancing between the CPUs and GPUs available
(see App.~\ref{sec:hybrid_approach}). These hybrid schemes should
be able to take advantage of the computational power available on these systems.

\section*{Acknowledgments}
This research was supported by the Fermi National Accelerator Laboratory (Fermilab),
a U.S. Department of Energy, Office of Science, HEP User Facility.
Fermilab is managed by Fermi Research Alliance, LLC (FRA),
acting under Contract No. DE--AC02--07CH11359.
MK and EB would like to thank Steffen Schumann for his support of the project, and the fruitful discussions leading towards it.

\appendix
\section{Proposal of a hybrid approach}
\label{sec:hybrid_approach}

\begin{figure}[t]
  \centering
  \includegraphics[width=0.8\textwidth]{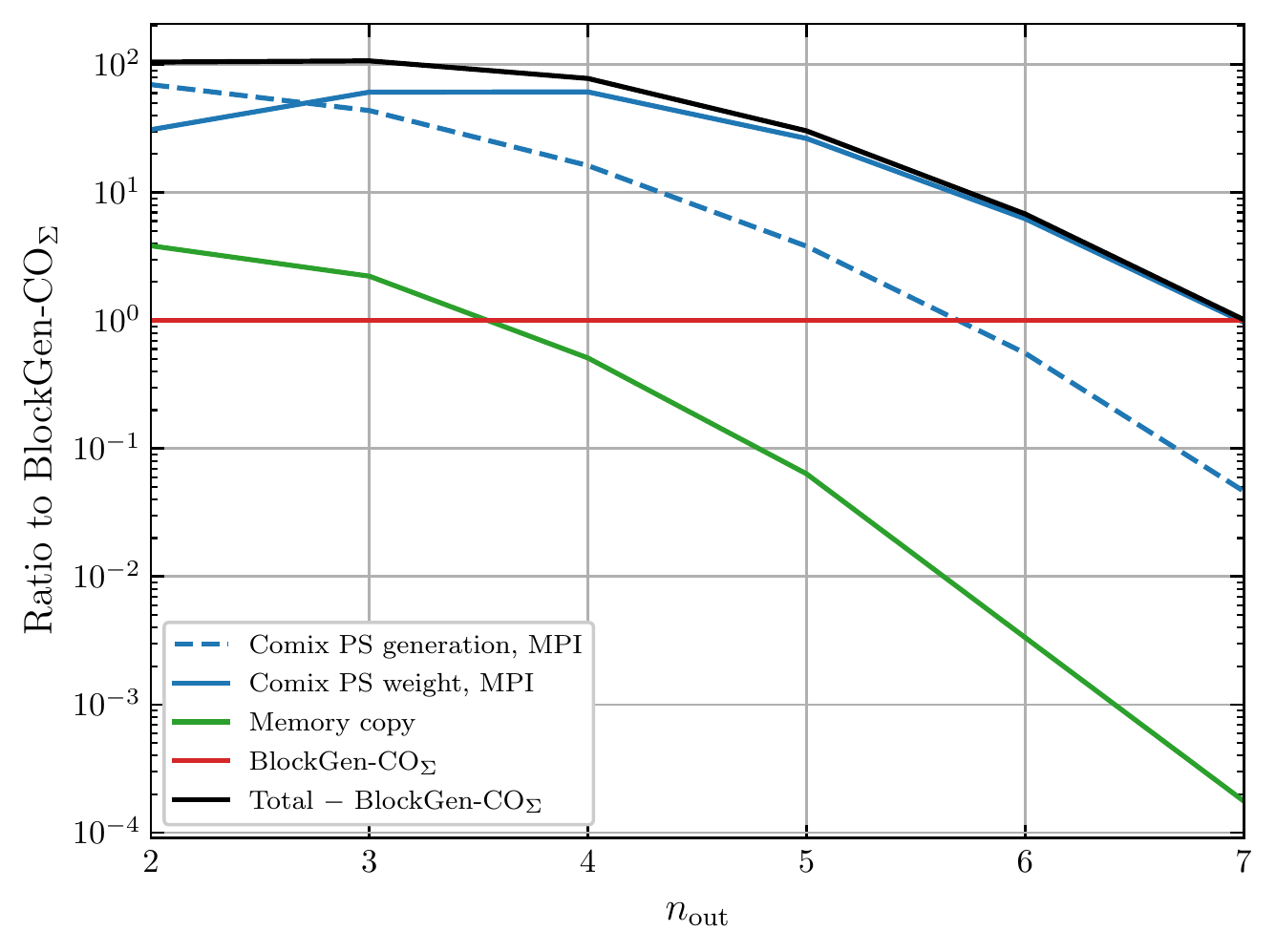}
  \caption{
    The timing ratio between phase-space point generation and
    weight computation using the recursive phase-space generator of Comix on the one hand,
    and the matrix element computation based on \newalgocosummed\ on the other hand.
    The ratio is also shown for the time required to copy the phase space points
    to GPU memory,
    and for the sum of everything except for \newalgocosummed.
    Comix timing is measured on an Intel$^\text{\textregistered}$
    Xeon$^\text{\textregistered}$ E5-2650 v2 8-core CPU, (\SI{2.60}{GHz}, \SI{20}{MB} cache),
    using MPI over 16 threads, while the GPU numbers are generated on a NVIDIA V100 (\SI{16}{GB} global memory, 5,120 CUDA cores, \SI{6144}{KB} L2 cache).}
  \label{fig:hybrid_proposal}
\end{figure}

In this appendix, we probe the possibility to construct a
hybrid event generator code, where some parts of the computation
are performed on the CPU, and other parts are performed on the GPU.
Such a division of tasks can involve
the phase-space generation, the calculation of phase-space weights,
and the calculation of the matrix elements.
In Fig.~\ref{fig:hybrid_proposal}, we show timing ratios
for these three tasks in comparison
to the color-summed matrix element calculation using \newalgocosummed.

We employ the recursive phase-space generator
of Comix~\cite{Gleisberg:2008fv} to provide the required time estimates
for the phase-space tasks.
This corresponds to the scenario where the CPU provides
the phase-space points and weights, while the GPU only evaluates
the matrix elements for those points.
In this scenario, a possible bottleneck might be the communication
between the CPU and the GPU.
We therefore also include a time estimate for copying the phase space
points to the GPU.
It turns out that the memory copy is not relevant for the overall timing
in this scenario.
Here we assume that the current MPI implementation of
Comix' phase-space generator would be changed to an OpenMP parallelized
algorithm, such that the memory copy to the GPU can take place within a single
process. 
The results in Fig.~\ref{fig:hybrid_proposal} show that such a division
of tasks would not be viable for multiplicities $n_\text{out} < 7$,
since the GPU would mostly wait for the CPU to generate phase space points,
weight them, and copy them over.

A more viable 
option for future development
might be a scheme where the CPU only generates phase space points,
while the GPU calculates both the associated weights
and the matrix elements.
In this scheme, the CPU would spend between about \SI{70}{\percent}
at $n_\text{out}=3$ and a few percent at $n_\text{out}=7$
of the time needed by the GPU
and would thus take over a sizable share of the overall workload
without forcing the GPU to be intermittently idle.
Only for $n_\text{out}=2$, the CPU time requirement would exceed
the one for the GPU, by about a factor of three.
Note that these estimates assume that the
phase-space weight calculation takes the same time on the GPU
as it does for Comix on the CPU.
If the GPU implementation leads to a sizable speed-up,
the GPU would need to wait for the CPU
even for $n_\text{out}\gtrsim 2$,
which might however still be an acceptable drawback
to accelerate the calculation for larger multiplicities.
It is beyond the scope of this work to implement a recursive GPU
phase-space generator needed to further assess this scenario.

\bibliography{refs.bib}

\nolinenumbers

\end{document}